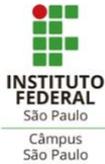

# Active Learning Methodology applied to a remote projectile launch experiment: students' first impressions

Carlos Antonio da Rocha[1], Matheus Santos Nogueira[2],

[1]Instituto Federal de Educação, Ciência e Tecnologia, Campus SP, Depto. de Ciências e Matemática,
carlosrocha@ifsp.edu.br

[2]Instituto Federal de Educação, Ciência e Tecnologia de São Paulo, Campus SP, Depto. de Construção Civil,
snogueiramatheus@gmail.com

## ABSTRACT

This study examines the impact of a remote laboratory experiment on Physics learning, using a case study approach. Societal advancements over the past century have spurred discussions regarding restructuring the current educational system, with active learning methodologies proposed as alternatives to conventional models. These active methods empower students to develop new skills and competencies beyond the academic knowledge imparted in lectures. To improve student learning outcomes in Physics, this research will develop an experimental model for application in engineering classes. This involves creating a laboratory experiment procedure for oblique projectile motion, constructing the necessary apparatus, recording the experiment on a film, presenting the video for a remote student's audience, and evaluating the model using questionnaires focusing on conceptual understanding and related questions.

## 1. INTRODUCTION

Alongside scientific and technological progress in the last century, a growing debate regarding the restructuring of the current educational system to meet the demands of a new reality has emerged (Miller; Shapiro; Hiding-Hamann, 2008). Active learning methodologies are presented as alternatives to the established conventional teaching approach. These approaches center the student's role in the learning process, empowering them to actively construct their own knowledge (Moreira, 2018; Henriques; Prado; Vieira, 2014).

Despite discussions of educational reform and the suggestion of improved teaching methods, traditional teaching methods, particularly in Physics, lag (Barbosa; Moura, 2013). Moreira (2017) notes that the current Physics curriculum often lacks relevance to modern concepts and technology, resulting in outdated and rote learning. Physics may be misinterpreted as a complete and static science. Moraes (2009) further highlights the lack of contextualization and practical applications connected to students' lived experiences as a source of confusion in physics education.

This study aims to analyze different active learning methodologies and assess the effect of a remote laboratory experiments on Physics learning. It will involve creating a physics laboratory experiment procedure, building the associated apparatus, and focusing on the concept of conservation of energy, specifically oblique projectile motion.

The purpose of this work is to discuss different active methodologies and the influence of the practical laboratory experiment on the learning process in physics classes, based on active methodologies and the physical concept of Oblique Launch, with the specific objectives being to:

a) Creating a script for a physics experiment, based on an extensive literature review, based on the concepts of active methodologies studied;
b) Construction of the didactic apparatus to be used, favoring low-cost materials and the use of cell phones and their applications for data analysis;
c) Evaluation of the material produced through remote experimentation with the engineering classes.

## 1.1 Active Learning Methodologies

Active learning methodologies are alternatives designed to replace outdated traditional teaching methods. Their key characteristics include student-centered learning, transforming the student into an active knowledge producer, and shifting the teacher's role to providing resources and support (Barbosa; Moura, 2013; Henriques; Prado; Vieira, 2014). Below are some active methodologies currently under discussion in academia.

### 1.1.1 Project-Based Learning (PBL)

Project-Based Learning (PBL) is an active learning approach where students learn by participating in the development of projects based on real-world situations. The goal is to provide a topic for investigation and delve deeper into theoretical academic concepts within practical contexts. PBL fosters critical thinking and problem-solving skills (Bender, 2014; Oliveira, 2019; Barbosa; Moura, 2013; Buck Institute for Education, 2008). Knowledge acquisition in PBL often involves studying and presenting an experimental setup within the school environment, although it is not limited to creating a physical product (Araujo, 2019; Oliveira, 2019).

Oliveira (2019) suggests that PBL should be structured around academic content and integrated into real-world student experiences. Larmer and Mergendoller (2015) outline eight key project characteristics, as seen in Figure 1: Challenging problem or Question; Sustained Inquiry; Authenticity; Student Voice and Choice; Reflection; Critical Analysis and Revision; Public Product; and Key Knowledge Understanding, and Success Skills.

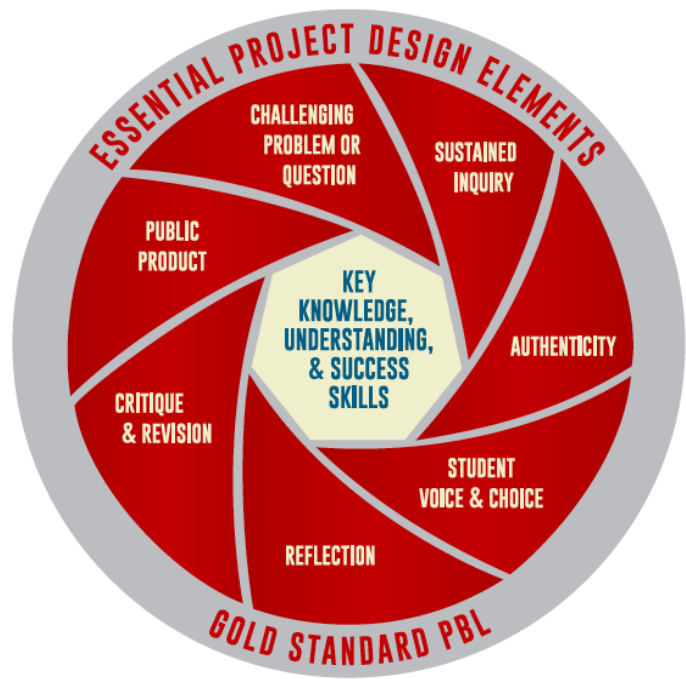

**Figure 1**: Key elements of Gold Standard PBL (Larmer; Mergendoller, 2015).

Araujo (2019) observes that in PBL, students collaborate to produce skills and knowledge by investigating the proposed problem and formulating hypotheses. Students are also able to observe the practical aspects of the studied content since projects are based on real-world scenarios.

The teacher's role in PBL is to provide the necessary resources for project development, guide the students, monitor progress, and address any questions (Araujo, 2019; Barbosa; Moura, 2013).

Evaluation in PBL is predominantly qualitative, as noted by Pasqualetto (2017), often involving records, grades, and questionnaires. Araujo (2019) and Oliveira (2019) also utilize records and questionnaires for data collection and analysis.

### 1.1.2 Task-Based Learning (TBL)

Task-Based Learning (TBL) introduces a task that stimulates self-directed learning (Barbosa; Moura, 2013; Paranhos et al., 2017). Studart (2019), Borochovicius and Tortella (2014) add that TBL is built upon students developing concepts, skills, and procedures by addressing real-world problems.

Barbosa and Moura (2013) emphasize that both students and teachers share the responsibility for learning. The teacher's role involves motivating students, providing guidance, and mediating discussions. The focus isn't on a final product, but on the investigative process and collaborative learning.

### 1.1.3 Science, Technology, Engineering, Art and Mathematics (STEAM) Education

STEAM education is a multidisciplinary pedagogical approach that has gained prominence in recent years. It integrates science, technology, engineering, art, and mathematics (hence, STEAM). It fosters integrated learning (Graça et al., 2020; Yakman; Lee, 2012).

For Yakman (2008), STEAM is considered a tool that contributes to an education capable of promoting problem-solving skills, creativity, critical thinking, collaboration and communication, without the academic content being affected. Figure 2 shows a diagram of STEAM's structure. It shows how the areas of knowledge are interconnected.

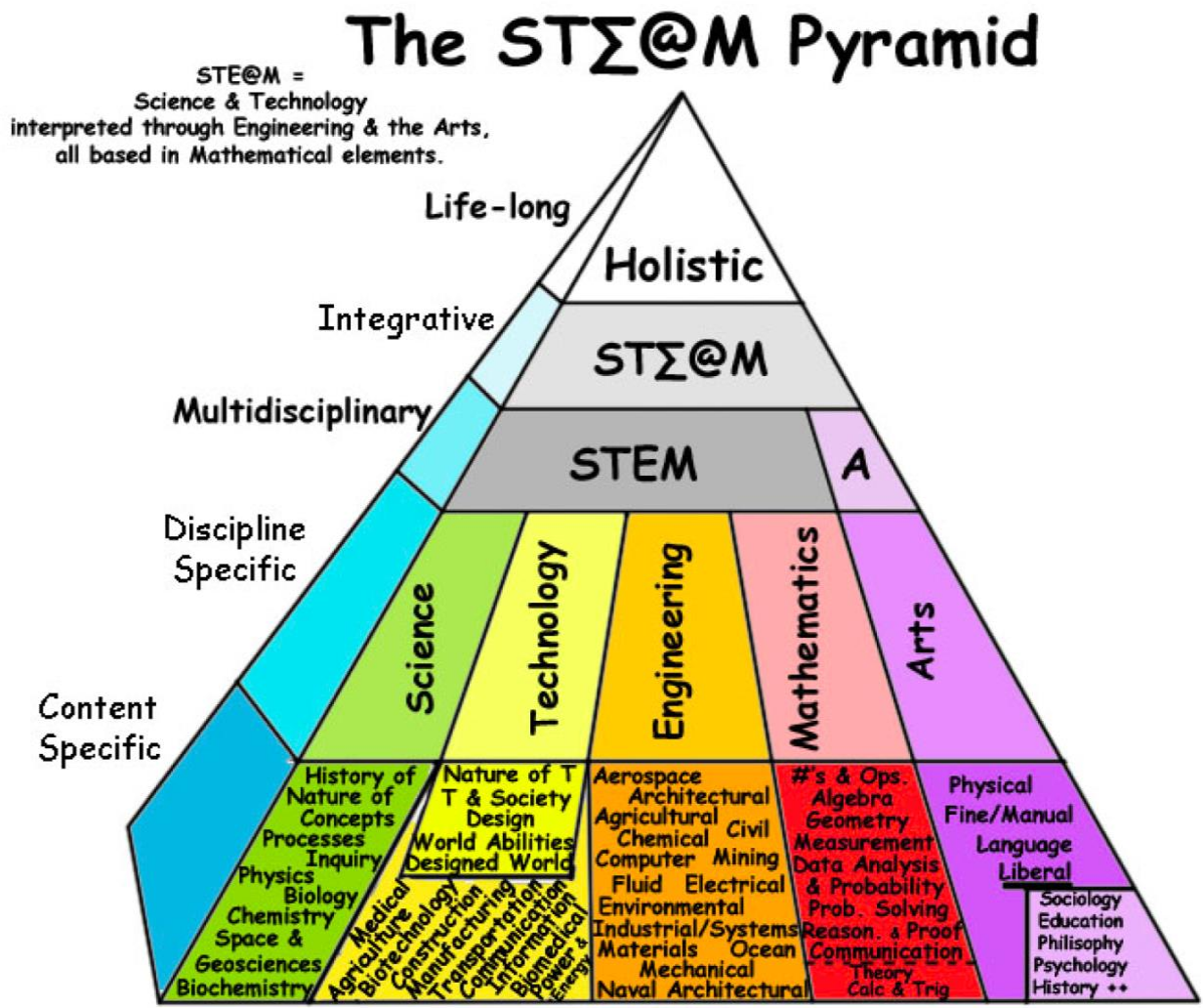

**Figure 2**: *G. Yakman diagram establishing a framework for giving structure to and analyzing the interactive nature of both the practice and study of the formal fields of science, technology, engineering, mathematics and the arts (Yakman, 2008).*

Kim and Chae (2016) cite creative design and emotional engagement as the foundation of STEAM, promoting self-directed learning and student interest.

While STEAM is still emerging in educational research, Bacich and Holanda (2020) point out the lack of clarity in its practical implementation. They argue that STEAM should be integrated into active learning approaches, especially PBL.

STEAM's structure aligns with PBL: it starts with a problem based on real-world scenarios, emphasizes investigative learning leading to an artifact, and culminates in the presentation of the project's results.

### 1.2 Projectile Motion (Ballistic Motion)

Projectile motion describes the movement of an object launched at an initial velocity ($v_0$), at an angle ($\alpha_0$) to the horizontal, subject to gravitational acceleration $\boldsymbol{a} = -g\boldsymbol{j}$. The resulting trajectory is a curve in a vertical plane known as a parabola (Halliday; Resnick; Walker, 2012; Sears et al., 2016). This model simplifies the analysis by neglecting air resistance and the Earth's curvature and rotation. A simplified coordinate system $(x, y)$ is used, where $x$ represents the horizontal axis and $y$ the vertical axis. Figure 3 shows the trajectory of a ballistic particle starting from the origin at a time $t = 0$.

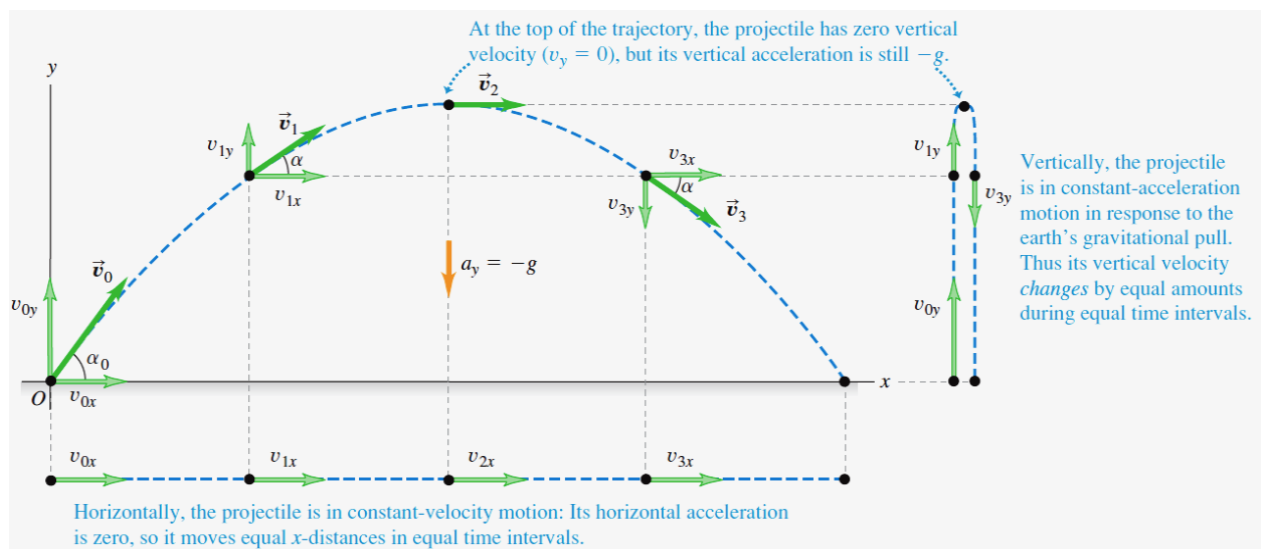

**Figure 3** - *Trajectory of a projectile leaving (or passing through) the origin at time t = 0. Extracted from (Sears et al, 2016).*

Horizontal motion is uniform (constant velocity) while vertical motion is uniformly accelerated due to gravity (Sears et al., 2016). Therefore, acceleration can be expressed as:

$$a_x = 0 \ e \ a_y = -g. \tag{1}$$

Note that $a_y$ is negative due to the chosen coordinate system direction.

Using the equations of kinematics (uniform motion and uniformly accelerated motion), we can determine the projectile's position and velocity at any given time. Equations 2-5 provide these calculations.

$$x(t) = x_0 + v_{0x} t \quad \rightarrow \quad x(t) = x_0 + (v_0 \cos \alpha_0) t \tag{2}$$

$$y(t) = y_0 + v_{0y} t - \frac{1}{2} gt^2 \quad \rightarrow \quad y(t) = y_0 + (v_0 \operatorname{sen} \alpha_0)t - \frac{1}{2} gt^2 \qquad (3)$$

$$v_x = v_{0x} \quad \rightarrow \quad v_x = v_0 \cos \alpha_0 \qquad (4)$$

$$v_y = v_{0y} - gt \quad \rightarrow \quad v_y = (v_0 \operatorname{sen} \alpha_0) - gt \qquad (5)$$

If the launch point is determined from the origin, then at $t = 0$ the initial positions on the axes will be zero, so $x_0 = 0$ e $y_0 = 0$.

From the projectile's position and velocity equations, other measures of movement can be calculated, such as the distance between the projectile and the origin, the scalar velocity and the direction of the projectile's velocity. The particle's position vector at each instant is given by Eq. (6); the distance between the projectile and the origin (r), which is the modulus of the position vector, is given by Eq. (7); the projectile's scalar velocity (v), at each instant, is given by Eq. (8) and the direction of the projectile's velocity, for each instant, is given by the angle $\alpha$ formed with the positive x-axis, shown in Eq. (9).

$$\boldsymbol{r}(t) = x(t)\boldsymbol{i} + y(t)\boldsymbol{j}. \qquad (6)$$

$$r = \sqrt{x^2 + y^2}. \qquad (7)$$

$$v = \sqrt{v_x^2 + v_y^2}, \qquad (8)$$

$$\tan \alpha = \frac{v_y}{v_x}. \qquad (9)$$

Beyond these Newtonian-based equations, projectile motion can also be analyzed through the conservation of mechanical energy. Conservative forces are those whose work does not depend on the path taken (Sears et al., 2016). Assuming an initial state (1) and a final state (2), the work of each force ($W_i$) must be added up, relating the initial and final kinetic energies (T) by the Eq. (10),

$$T_1 + \sum_i W_i = T_2 ; \qquad (10)$$

which results in the conservation of mechanical energy, Eq. (11)

$$E_{m_1} = E_{m_2} ; \qquad (11)$$

where $E_m$, the mechanical energy, represents the sum of the kinetic and potential energies, assuming no friction force acting.

Knowing further that the kinetic and potential energies are given by the formulas in Eq. (12)

$$V_g = Wy \quad ; \quad V_e = \frac{1}{2} k s^2 \quad ; \quad E_c = \frac{mv^2}{2}, \qquad (12)$$

this allows us to determine the projectile's velocity at a specific point, its maximum height, and the spring constant of the launch mechanism.

## 2. METHODOLOGY

This study employs a research methodology involving a literature review of active learning techniques and the development of an experimental classroom model using a custom-built apparatus for projectile motion. The goal is to investigate the impact of a remote approach on an experimental Physics lab instruction. The project focus on creating an inexpensive and easily replicable device for demonstrating Physics concepts within an undergraduate engineering classroom setting.

The experimental model incorporates conceptual questions on oblique projectile motion and the practical application of the concept. The technology used for data collection and analysis included standard computer and mobile devices. Software employed included Microsoft Excel for data organization, OriginLab for statistical analysis, and Tracker for video analysis. These programs are readily accessible and commonly used by engineering students. All project documents and materials are available via a shared Google Drive folder[1].

### 2.1 Apparatus Construction

To study the impact of remote activities on Physics learning, considering the emphasis on practical elements in many active learning methods, a projectile launcher was constructed for use in physics classes focusing on oblique projectile motion. This launcher utilizes readily available, low-cost materials and operates via spring compression. Two launcher designs were evaluated before settling on a final model.

The initial design comprised a wooden frame resting on a pine wood base, with a metal rod, spring, and projectile launch mechanism. The frame was attached to the base with a pivot allowing for adjustment of the launch angle, according to Figure 4.

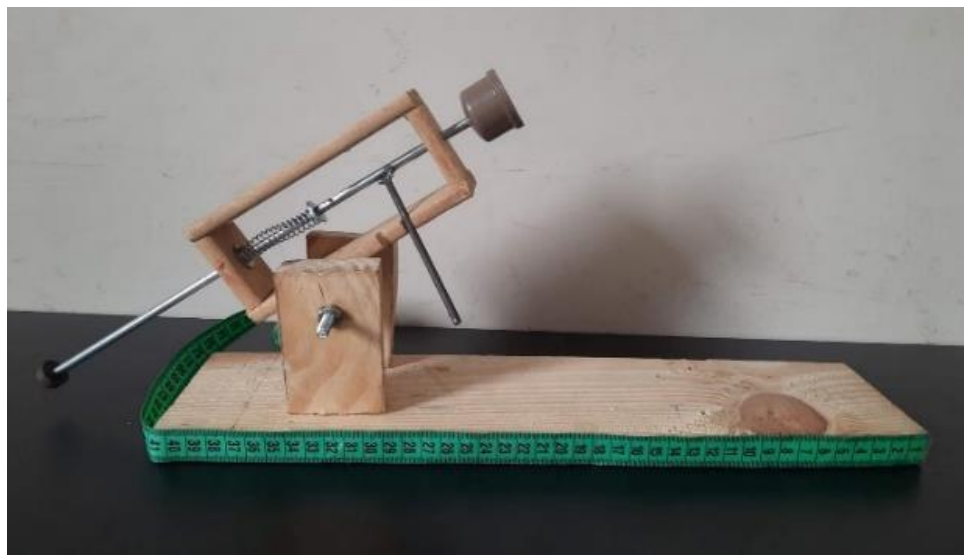

*Figure 4* – *Starting model of the projectile launcher.*

---

[1] Link to access the shared folder: drive.com/shared_folder

Technical report presented in December 2021 for the conclusion of a Scientific Initiation research project.

This initial model had flaws in its locking and angle adjustment mechanisms, affecting data acquisition and introducing errors into projectile motion analysis. A revised model was created using a different support structure and angle adjustment system. The second model consists of a box-shaped wooden structure (plywood) mounted on a support base, with a metal rod, spring, and launch mechanism, as illustrated in Figure 5.

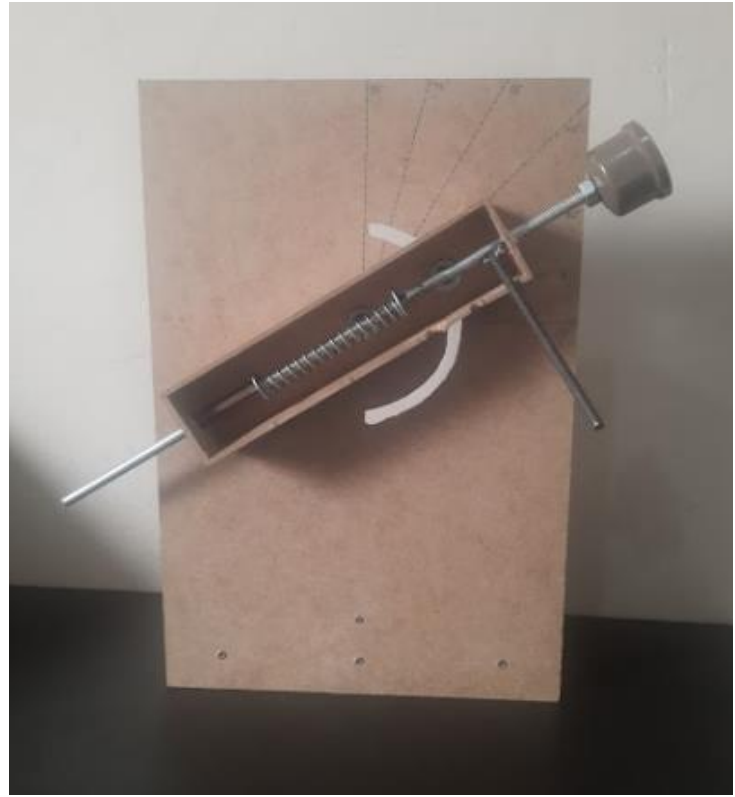

*Figure 5 – Final model adopted for the projectile launcher.*

The main improvement involved the support and angle adjustment mechanisms. The support uses a weighted box at the rear, and the angle is adjusted using a pin and a cutout arc on the plywood frame. The wooden frame has two holes for locking the metal rod, as shown in Figure 6.

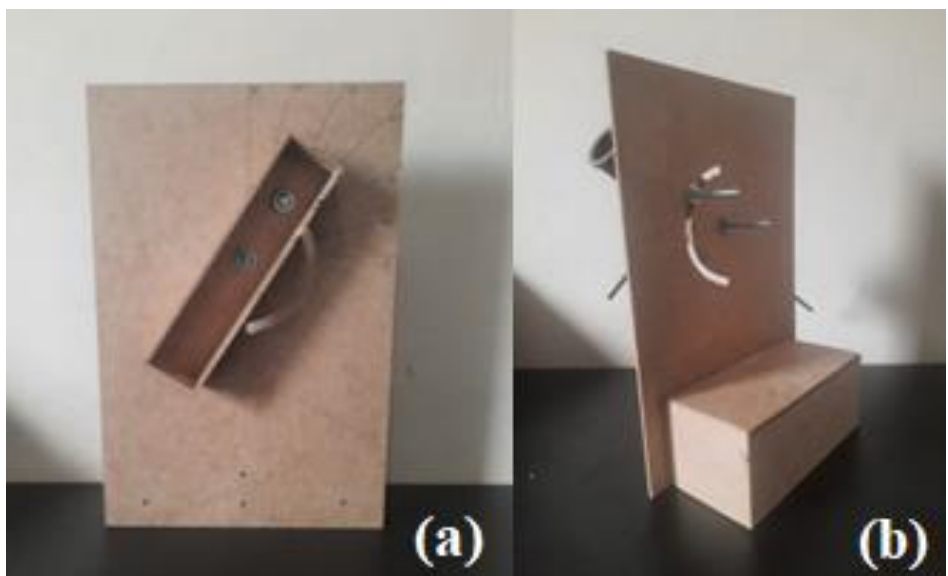

*Figure 6 – Wooden frame, showing (a) frontal view and (b) rear view*

Technical report presented in December 2021 for the conclusion of a Scientific Initiation research project.

The metal rod includes a spring held in place by washers and a cylindrical locking mechanism. It also has a projectile launch nozzle and a perpendicular locking rod, seen in Figure 7.

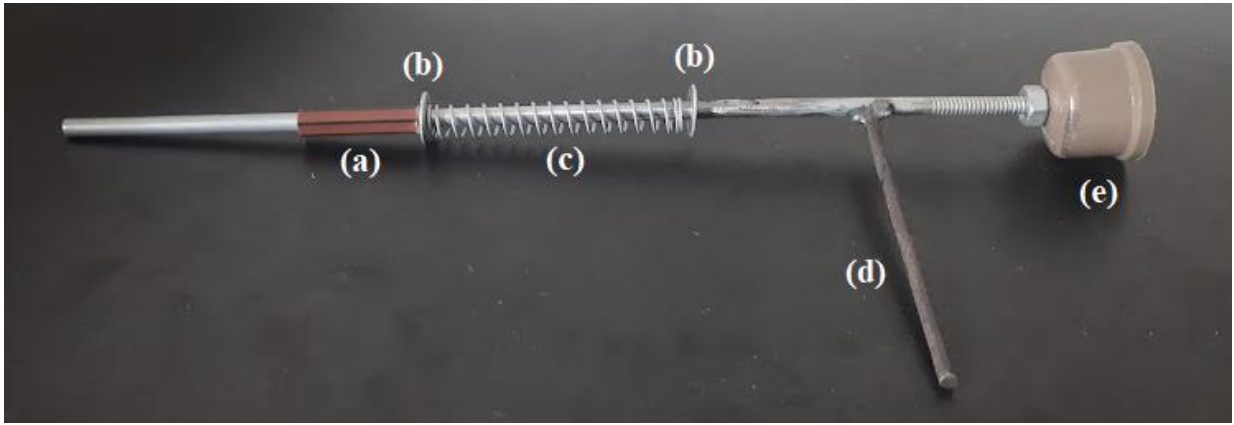

**Figure 7** - *Metal rod. (a) Cylindrical lock (b) Washers (c) Spring (d) Locking rod (e) Nozzle.*

## 2.2 Launcher Operation

The launcher operates by compressing and releasing the spring attached to the metal rod. First, the metal rod is moved to compress the spring, then locked into place within the plywood frame. A projectile is inserted into the launch nozzle, and the launch angle is adjusted. The locking mechanism is then released, allowing the spring to return the system to its initial position and launch the projectile in an oblique trajectory. The system is illustrated in Figure 8.

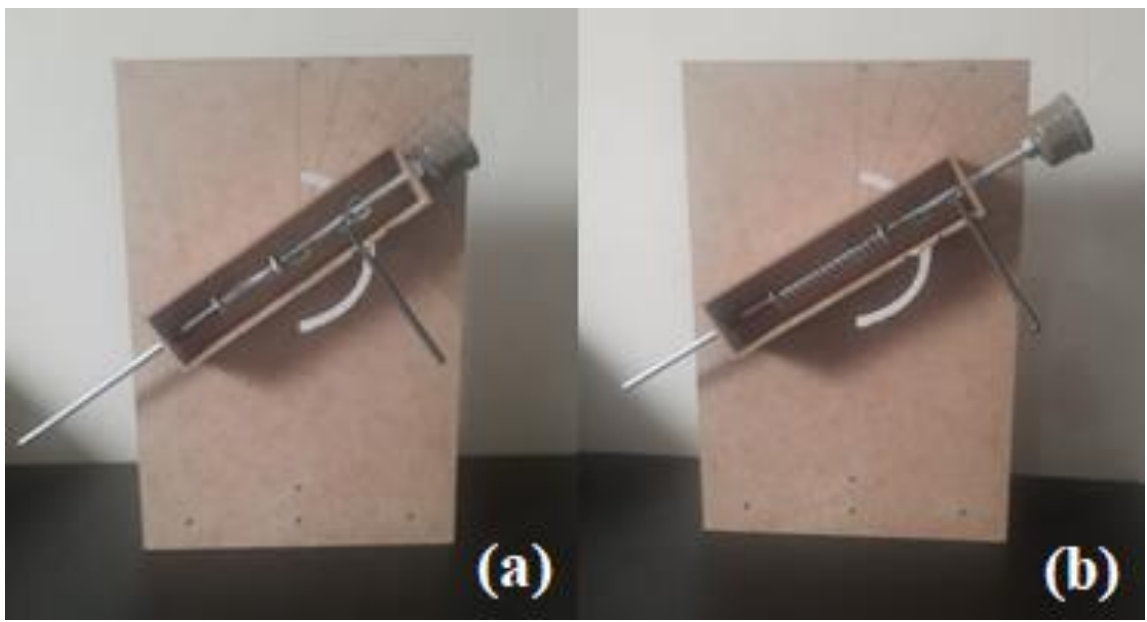

**Figure 8** - *Launcher operation (a) Pre-launch (spring compressed) (b) Post-launch (spring released).*

## 3. DATA PROCESSING

Oblique projectile motion was analyzed using video recordings of a projectile of known mass launched at predetermined angles. Three software packages were employed for data acquisition and processing: Microsoft Excel for spreadsheet management, OriginLab for statistical and graphical analysis, and Tracker for video analysis and modeling.

Videos were recorded using smartphones at 30 frames per second (fps) at normal playback speed. The recording environment was controlled to minimize external factors such as wind and vibrations. A white background with metric references was used to aid in motion analysis, as painted in Figure 9.

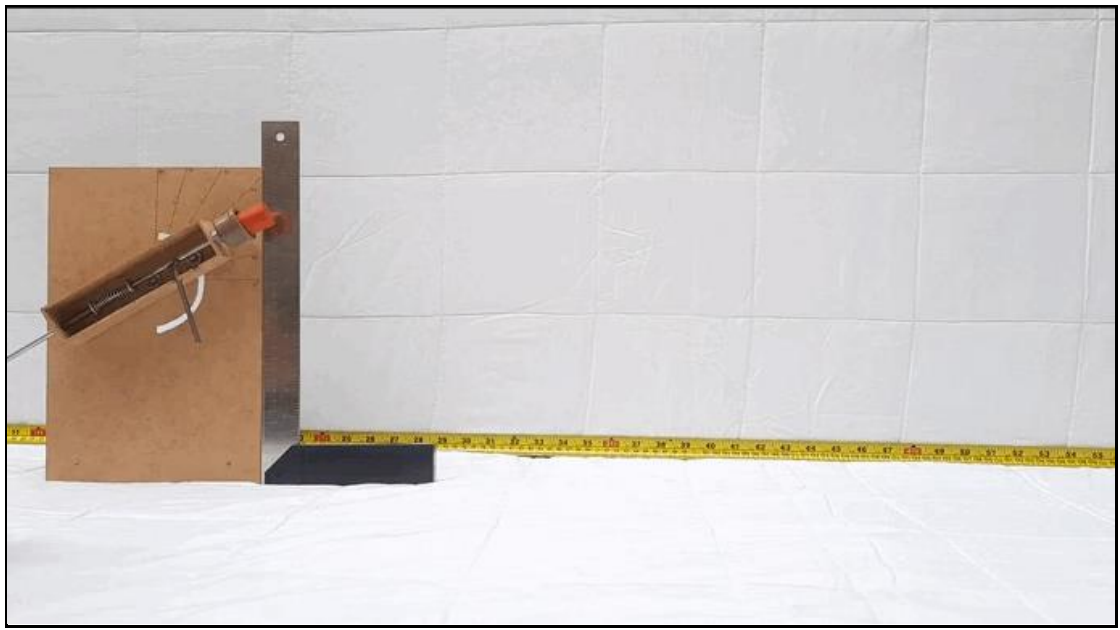

*Figure 9* - *Frames and backdrops used to record the launch videos.*

The videos were imported into Tracker software, where the "automatic trajectory" feature was used to map the projectile's path. This provided the experimental position and time data for the trajectory. Note that the origin and measurement points were defined using the known metric references. The nozzle's position at the launch moment served as the origin for the Cartesian coordinate system, as seen in Figure 10.

The video analysis focused on the time interval from spring release (initial position) to the projectile hitting the ground (final position). Figure 10 displays the Tracker software interface showing the projectile trajectory.

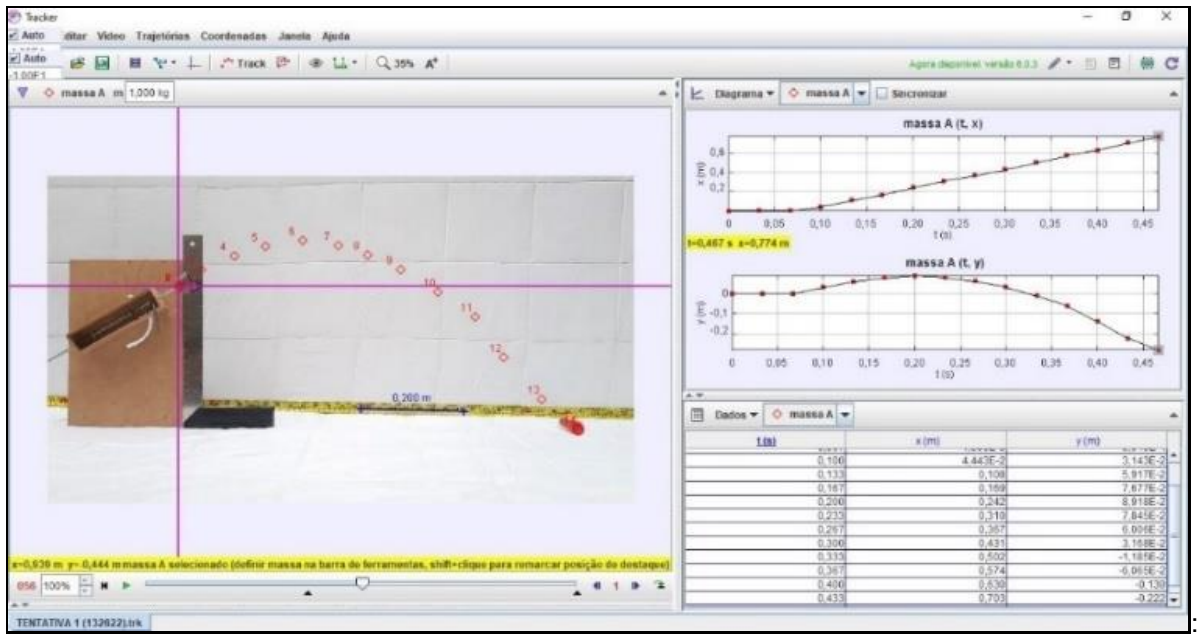

**Figure 10** - Tracker software screen for video analysis.

The experimental data was then used to plot parabolic curves representing the projectile trajectory in OriginLab software. These curves show *y* as a function of *x* (projectile position on the *y*-axis relative to the *x*-axis) and *y* as a function of *t* (projectile position on the y-axis relative to time). OriginLab also generated corresponding equations for each curve, declared in Figure 11.

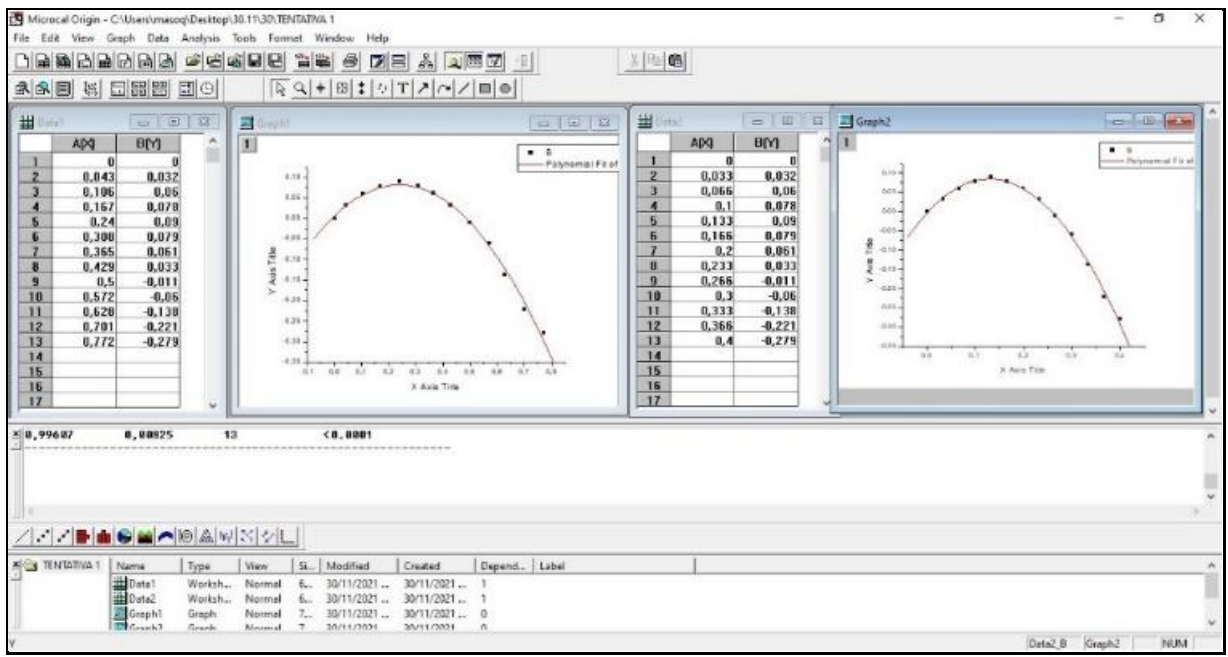

**Figure 11** - Origin Lab software screen for analyzing the experimental points.

Using the equations of projectile motion along the *x* and *y* axes (Equations 2 and 3), along with equations derived from the OriginLab curve fits, experimental values for initial velocity, gravitational acceleration, and launch angle were calculated. The conservation of energy principle was then applied to determine energy loss during the launch (Equations 13-15). These calculations were performed and organized using Microsoft Excel and online spreadsheet software.

Knowing that the 2nd grade polynomial equations gotten from OriginLab software are written as

$$y = A + B1 \cdot x + B2 \cdot x^2; \tag{13}$$

we got, for the *y = f(x)* curve, the fitted Eq. (14) and, for the *y = f(t)* function, the fitted Eq. (15).

$$y = A + B1 \cdot x + B2 \cdot x^2 \rightarrow y(x) = y_0 + \text{tg}\,\theta \cdot x + \frac{(-g)}{2v_0^2 \cos^2\theta} \cdot x^2 \tag{14}$$

$$y = A + B1 \cdot t + B2 \cdot t^2 \rightarrow y(t) = y_0 + v_0 \text{sen}\,\theta \cdot t + \frac{(-g)}{2} \cdot t^2 \tag{15}$$

Finally, knowing the terms of the equations obtained through Origin Lab, it was possible to calculate the experimental values of initial velocity, acceleration due to gravity and launch angle. And, using the theoretical concepts of energy conservation, determine the loss of energy in the system during the launch.

## 4. REPLICABILITY AND 3D PRINTING

Given the student-centered nature of active learning methodologies and the importance of project-based activities in promoting student engagement, expanding hands-on experimental opportunities and broadening the scope of knowledge production are highly desirable. Therefore, in addition to the wooden launchers, two 3D-printable prototype designs were created using AutoCAD and SolidWorks software for 2D and 3D modeling.

These new models aim to enhance replicability (standardization simplifies production), encourage student participation in construction, and promote interdisciplinarity (integrating programming and construction). This allows students to develop their programming skills while building and using the launcher, if they want to.

The first prototype is box-shaped, like the wooden launchers; the second is cannon-shaped. Both use locking mechanisms and operate by spring compression, as shown in Figure 12. The 3D models are in .SLDPRT format and require conversion to a format compatible with the chosen 3D printer. These files are available on the Google Drive folder mentioned earlier.

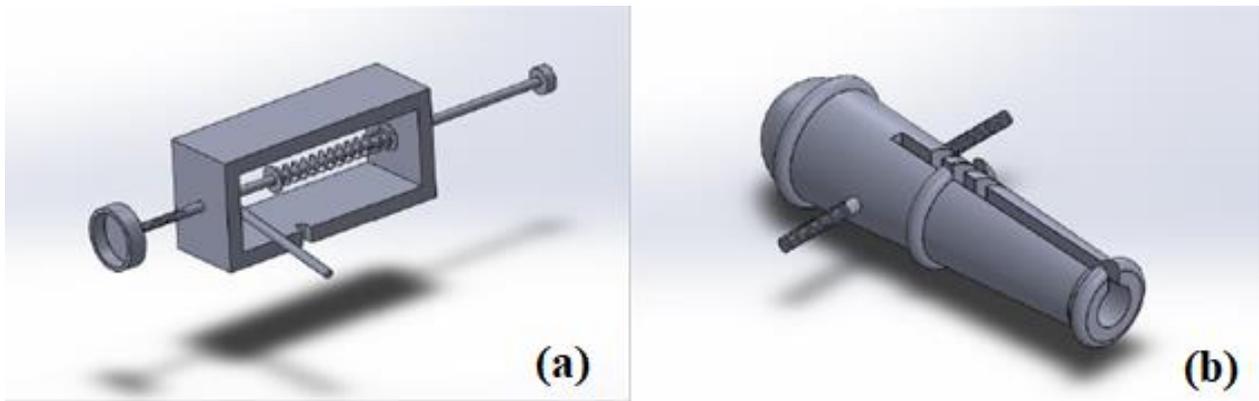

Figure 12 - Launchers in 3D (a) Box-shaped launcher (b) Cannon-shaped launcher

It should be emphasized that the suggestion of 3D printing is based on the availability of equipment and laboratories in the local area. These resources include an internal laboratory at the Federal Institute of Education, Science, and Technology of São Paulo (IFSP) and access to publicly available 3D printing facilities in São Paulo (FAB LAB LIVRE SP), or commercial printing services.

## 5. RESULTS AND DISCUSSION

To apply the studied active learning methodologies and assess the impact of remote activities on knowledge acquisition, an experimental procedure was implemented in two engineering classes: civil and mechanical engineering. A questionnaire was used for assessment, including seven questions on oblique projectile motion and conservation of energy, and seven conceptual questions on experimental-based teaching methodologies.

The filmed experiment was presented for two classes of 1st year Engineering. Due to time constraints imposed by the academic calendar, a simplified version of the procedure was provided to the civil engineering class, with data analysis equations and software instructions (Tracker and OriginLab) made available. Software use was optional but encouraged, with guidance provided. The mechanical engineering class was required to use the software. The civil engineering class yielded 26 responses, and the mechanical engineering class yielded 11.

The exercise involved analyzing a video of a 0.030 kg projectile launched at a 30° angle. Students were given the spring constant (666.3 N/m), initial spring compression (0.034 m), and the mass of the spring-projectile system (0.108 kg). They were then asked questions about oblique projectile motion.

The first question focused on conservation of mechanical energy, asking students to calculate the stored elastic potential energy in the spring to determine the launch velocity and the initial mechanical energy. Civil engineering students answered this question correctly. Mechanical engineering students, despite using the same approach, showed discrepancies in parts (b) and (c), leading to subsequent errors. This stemmed from

Technical report presented in December 2021 for the conclusion of a Scientific Initiation research project.

incorrectly using the mass of the projectile alone in calculations instead of the combined system mass (as specified in the problem statement). While calculation methods were accurate in both classes, error propagation analysis was not performed. Results were generally consistent with expected values, and an example is shown in Figure 13).

```
1) K = 666,3 N/m ;   X = 0,0340 m ;   m_r = 0,108 Kg ;   m_p = 0,030 Kg
a) Epe = k·x²/2  →  666,3 · 0,0340²/2 = 0,3851 J
b) Ec = Epe →  0,108/2 v² = 0,3851 → V = 2,6705 m/s
c) Ec = M·v²/2  →  0,03 · 2,6705²/2 = 0,1070 J → E_minicia = 0,1070 J
```

Figure 13 - Resolution of question 1 produced by one of the students assessed.

The second question asked students to apply theoretical equations to experimentally determined values, justifying the results obtained. Most students responded correctly with experimental values approximating theoretical values. An example is shown in Figure 14.

```
Questão 2
Y(t) = Y₀ + V₀·senθ·t − g/2·t²  ;  Y(t) = A + B₁·t + B₂·t²

A = Y₀ = −0,003 m ;  B₁ = V₀·senθ = 1,351 m/s ;  B₂ = −g/2 = −5,196
                                                     g = 10,392 m/s²

Portanto, tendo em vista que os valores desejáveis eram Y₀ = 0 m e
g = 9,81 m/s², temos que os valores do Origin lab são sim razoáveis
```

Figure 14 - Resolution of question 2 by one of the students assessed. The written comment at the end is "Therefore, since the desired values were $y_0 = 0$ and $g = 9{,}81$ m/s², the values provided by the fitting curve were acceptable."

The third question, like the second, required calculations of additional parameters (e.g., launch angle), verifying the compatibility of experimental and theoretical results. Figure 15 presents an example response.

```
3-   0,6015 = tan A = 31°              Os valores são
     V₀ sen 31°: 1,419                  razoáveis com o
     V₀ = 1,419 → V₀ = 2,756 m/s        que seria esperado
           0,515
     y₀ = 0
```

Figure 15 - Resolution of question 3 produced by one of the students assessed.

Technical report presented in December 2021 for the conclusion of a Scientific Initiation research project.

The fourth question compared the calculated velocities from questions 1 and 4. The discrepancies in question 1 due to incorrect mass usage resulted in approximately 80% difference for those students. Students who used the correct mass showed a much smaller difference of approximately 6%, as shown in Figure 16.

Figure 16 - Resolution of question 4 produced by one of the students assessed.

The fifth question verified the consistency of calculated values with given flight time (0.4 seconds) and range (0.772 meters). Most students answered correctly, although some lacked detailed explanations, as seen in an example in Figure 17.

Figure 17 - Resolution of question 5 produced by one of the students assessed.

The sixth question, again related to energy conservation, asked students to determine the final potential energy of the projectile at impact. Most responses were correct. An example is shown in Figure 18.

Figure 18 - Resolution of question 6 produced by one of the students assessed.

Technical report presented in December 2021 for the conclusion of a Scientific Initiation research project.

The seventh and final question involved calculating the final velocity and kinetic and mechanical energies at impact. Most students answered correctly, but some errors arose due to propagation of previous calculation errors. An example is given in Figure 19.

Figure 19 - Resolution of question 7 produced by one of the students assessed.

Beyond the physics questions, students were also asked about the proposed model and the integration of a remote experiment in the laboratory work. In response to a conceptual question, students reported that the suspension of laboratory activities during the pandemic negatively impacted their learning, even with remote experiments. However, they acknowledged that in-person lab work significantly enhances understanding. An example answer is shown in Figure 20.

Figure 20 - Resolution of conceptual question 1 produced by one of the students assessed. The text says: "Certainly. Even though I carried out this and other experiments online, I believe that I was able to understand the content given during this semester. However, the hands-on experience is more efficient for the way I learn."

The second conceptual question explored the model's contribution to learning. Students generally felt that the model aids understanding, especially when combined with theory. The practical aspect facilitates visualization and comprehension. An example of answer is shown in Figure 21.

**Figure 21** - *Resolution of conceptual question 2 produced by one of the students assessed. The text says: "Having practical classes on one day and theoretical classes on another contributes to diversity in the way teaching is done, making it easier to assimilate the content by analyzing theoretical content in practice."*

The third conceptual question asked students to evaluate the model's strengths and weaknesses. Positive feedback focused on improved understanding, enhanced motivation, and the opportunity for self-directed learning. Negative feedback noted the time required, resource limitations, and potential video quality issues, as noted in one answer given in Figure 22.

[Handwritten text in image]

**Figure 22** - *Resolution of conceptual question 3 produced by one of the students assessed. The text says: "Positive: better understanding, improved results and objectivity. Negatives: online format, no personal contact between teacher and student."*

Students viewed the model positively in the fourth conceptual question, finding that it addressed their learning needs without requiring additional resources, as seen in Figure 23.

[Handwritten text in image]

**Figure 23** - *Resolution of conceptual question 4 produced by one of the students assessed. The written text says: "Yes, it helped a lot to better understand how the oblique launch works."*

In the fifth conceptual question, students provided examples of oblique projectile motion from various sports and activities. Examples are given in Figure 24.

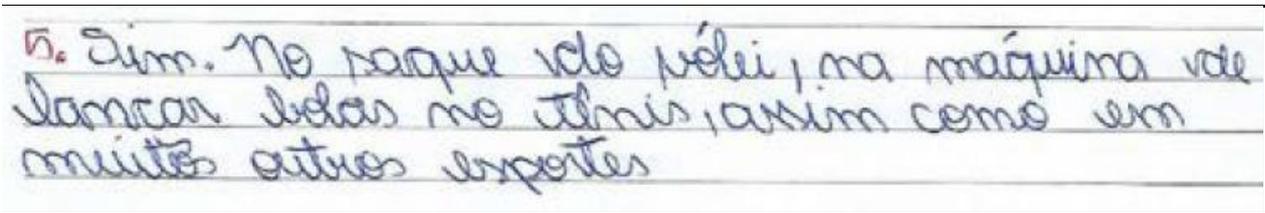

*Figure 24* - Resolution of conceptual question 5 produced by one of the students assessed. The written text says: "Yes, in the volleyball serve, in the ball-throwing machine in tennis, as well as in many other sports."

The sixth question explored the value of building a similar device. Students highlighted benefits to learning and the development of additional skills. Figure 25 shows an answer.

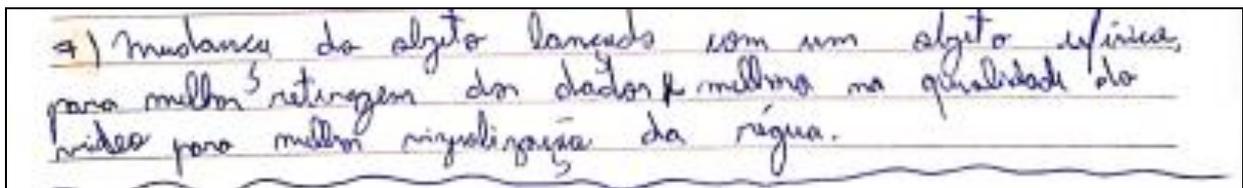

*Figure 25* - Resolution of conceptual question 6 produced by one of the students assessed. The answer says: "Yes. As I suggested earlier, this proposal to build a device and analyze all the data is very interesting in terms of more practical and less monotonous learning. It could also be applied to other concepts, such as free fall, for example."

Finally, in the seventh conceptual question, several suggestions were given focused on improvements to the support structure, locking mechanism, projectile design, and video recording quality for future iterations. Figure 26 shows one answer.

*Figure 26* - Resolution of conceptual question 6 produced by one of the students assessed. The text says: "Change the object launched to a spherical object in order to obtain better data. Improved video quality to better visualize the ruler."

## 6. CONCLUSION

The results show that the model used in a remote activity significantly aided in knowledge acquisition and motivation. While some students did not answer all physics questions perfectly, they demonstrated a solid grasp of the concepts, capable of correctly formulating hypotheses and applying mathematical equations.

Active learning methodologies (particularly project-based approaches involving apparatus construction) are effective for promoting student engagement and the development of valuable skills. Students indicated that they appreciated the practical element, preferring in-person activities.

Technical report presented in December 2021 for the conclusion of a Scientific Initiation research project.

It's important to note the limitations imposed by the pandemic, which required a remote and shortened experimental implementation. Factors like class size, teacher training in active learning, and resource availability should be considered when adopting this approach.

Ultimately, this study confirms the good job played by remote experiments but makes clear the vital role of hands-on laboratory activities in engineering education. Active methods enhance understanding, especially in Physics, by allowing students to visualize and reinforce theoretical concepts. The pandemic's restrictions highlighted the value of these experiences for student learning. The ongoing discussion about educational reform, exemplified by the implementation of the "Novo Ensino Médio" in Brazil, underscores the growing need for approaches that foster skill development for students' success in society.